\documentclass[12pt]{iopart}

\usepackage{iopams} 
\usepackage{bm,graphics,graphicx,epsfig,color}

\begin{document}

\title[PN effects in hierarchical triples]{Post-Newtonian effects in $N$-body dynamics: Relativistic precession and conserved quantities in hierarchical triple systems}

\author{Clifford M. Will$^{1,2}$}

\ead{cmw@physics.ufl.edu}
\address{
$^1$ Department of Physics,
University of Florida, Gainesville FL 32611, USA
 \\
$^2$ GReCO, Institut d'Astrophysique de Paris, UMR 7095-CNRS,
Universit\'e Pierre et Marie Curie, 98$^{bis}$ Bd. Arago, 75014 Paris, France
}

\begin{abstract}
Conventional approaches to incorporating general relativistic effects into the dynamics of $N$-body systems containing central black holes, or of hierarchical triple systems with a relativistic inner binary, may not be adequate when the goal is to study the evolution of the system over a timescale related to relativistic secular effects, such as the  precession of the pericenter. 
For such problems, it may necessary to include post-Newtonian ``cross terms'' in the equations of motion in order to capture relativistic effects consistently over the long timescales.  Cross terms are post-Newtonian (PN) terms that explicitly couple the two-body relativistic perturbations with the Newtonian perturbations due to other bodies in the system.
In this paper, we show that the total energy and the normal component of total angular momentum of a hierarchical triple system is manifestly conserved to Newtonian order over the relativistic pericenter precession timescale of the inner binary {\em if and only if} PN cross-term effects in the equations of motion are taken carefully into account.  
\end{abstract}


\maketitle

\section{Introduction and summary}
\label{sec:intro}

The study of general relativistic effects in orbital dynamics has evolved in recent years well beyond the simple two-body problem that was of such historic importance for the theory.   Galactic star clusters with massive central black holes~\cite{2006ApJ...645.1152H,mamw2}, triple systems with a relativistic compact inner binary~\cite{2002ApJ...576..894M,2002ApJ...578..775B,2003ApJ...598..419W,2011MNRAS.411..565M,2013PhRvL.111f1106S,2013ApJ...773..187N}, binary black hole coalescence in the presence of a third body~\cite{2011PhRvD..83h4013G,2011PhRvD..84j4038G,2014MNRAS.439.1079A}, and even the stability of the solar system~\cite{2008CeMDA.101..289B} have been studied using combinations of $N$-body techniques and relativistic dynamics.

In a recent paper~\cite{2014PhRvD..89d4043W} (hereafter referred to as Paper I), we argued that conventional approaches to incorporating relativistic effects in such analyses may not be adequate when the goal is to study the evolution of the system over a timescale related to relativistic secular effects, notably the  precession of the pericenter.   In the conventional approach, one typically augments the $N$-body Newtonian equations of motion by two-body post-Newtonian (PN) relativistic corrections, where the two bodies in question might be a chosen star and the central black hole in the galactic core problem, or the tight binary system in the hierarchical three-body, or Kozai-Lidov problem.   We argued that, for such problems, it may necessary to include post-Newtonian ``cross terms'' in the equations of motion in order to capture the relativistic effects consistently over the long timescales.

In the context of a hierarchical triple system, the idea of cross terms is as follows: a relativistic effect in the inner binary, such as the pericenter advance, is proportional to $Gm/ac^2$, where $m$ and $a$ are the mass and semimajor axis of the binary, and $G$ and $c$ are the gravitational constant and speed of light, respectively.   A Newtonian effect due to the third body at a distance $R \gg a$ is proportional to $(m_3/m)(a/R)^n$, where $n$ is a power depending on the degree to which the field of the third body is expressed in a multipole expansion ($n = 3$ corresponds to the leading quadrupole order).   A PN effect due to ``cross terms''  would be proportional to $(Gm/ac^2) \times (m_3/m)(a/R)^n$.  On the face of it, this is a smaller effect than either the pure PN effect or the third-body effect, when $(a/R)^n \ll 1$.  However, if it is a secular effect, {\em and} if one is interested in how this effect grows over a relativistic timescale induced in the binary, $T_R \sim T_B (ac^2/Gm)$, where $T_B$ is the binary period, then the effect could be ``boosted'' from a PN-level effect to a Newtonian-level effect.  This could have hitherto unforeseen consequences in long-term evolutions of such systems.   
In the context of stellar clusters with a central black hole, $m$ becomes the mass of the central black hole, $a$ becomes the semimajor axis of a chosen star $b$, $m_3$ becomes $m_c$ and $R$ becomes $R_{bc}$, summed over the other stars in the cluster. 

The origin of these ideas was the simple two-body problem described in Paper I of a test particle moving in the gravitational field of a body with mass $M$ and quadrupole moment $Q_2$, including the standard PN corrections from the Schwarzschild part of the metric, whose main consequence is the advance of the pericenter.  The Newtonian conserved energy per unit mass, evaluated at pericenter of the orbit is given by 
\begin{equation}
E = - \frac{GM}{2a} + \frac{GQ_2}{2}\left ( \frac{1+e}{p} \right )^3 \bigl (1-3\sin^2\iota\sin^2 \omega  \bigr) 
\,,
\label{EnewtQ}
\end{equation}
where $a$, $e$, $\iota$ and $\omega$ are the osculating semimajor axis, eccentricity, inclination and pericenter angle of the orbit, and $p = a(1-e^2)$ (for a pedagogical introduction to osculating orbit elements see~\cite{PW2014}).  To this order, $a$, $e$ and $\iota$ do not experience secular changes, but $\omega$ grows linearly with time at the rate $6\pi GM/c^2p$ because of relativistic effects due to the mass $M$ (assumed to dominate over the pericenter advance induced by the quadrupole moment).  When $\omega$ changes by a macroscopic amount, say $\pi/2$, the energy apparently changes by a Newtonian-order amount, in violation of the basic conservation law.  We showed in Paper I that when cross terms of order $(GM/ac^2) \times  (Q_2/Ma^2)$ were included in the post-Newtonian equations of motion, and when the equations for the perturbed orbit elements were carefully solved (including internally generated cross-term contributions), the semimajor axis $a$ suffered a secular change per orbit that was of PN order and also depended on $Q_2$, which, when integrated over a pericenter precession timescale, was boosted to a Newtonian-order variation in $a$ that {\em exactly} compensated for the $\omega$ dependence in Eq.~(\ref{EnewtQ}), leaving an energy expression that was manifestly constant over a pericenter precession timescale.   

This unusual result motivated us to suggest that such cross-terms should be taken into account in other contexts, such as stellar clusters with central massive black holes and hierarchical triple systems.  Accordingly, in Paper I we wrote down the truncated post-Newtonian equations of motion, including all relevant cross terms, in a ready-to-use form either for numerical $N$-body simulations of clusters with a central black hole or for studies involving perturbations of orbit elements in hierarchical triple systems.  For the simple case of a hierarchical triple with the third body in a circular orbit, we solved the osculating orbit element perturbation equations for the binary explicitly, including the cross-term effects.  In this paper, we shall apply those results to demonstrate explicitly that the total energy $E$ and the component of angular momentum of the system perpendicular to the orbital plane of the third body $L_Z$ are conserved over a pericenter precession timescale of the inner binary if and only if the PN cross term effects are included.

In Sec.\ \ref{basic}, we review the basic equations of Paper I for hierarchical triple systems, and in Sec.\ \ref{conservation} we derive the conserved energy $E$ and total angular momentum ${\bm L}$ for the system, to PN cross-term order, and show that the lowest-order expressions (Newtonian plus post-Newtonian) are apparently not conserved over a pericenter precession timescale, presenting the same conundrum as in the quadrupole problem.  In Sec.\ \ref{resolution} we show that incorporating the full secular evolution of the orbit elements including PN cross terms completely resolves the conundrum.  Concluding remarks are presented in Sec.\ \ref{conclusion}.

\section{PN equations of motion for hierarchical triple systems}
\label{basic}

We consider a three-body system in which two bodies of mass $m_1$ and $m_2$ are in a close orbit with separation $r$, and a third body of mass $m_3$ is in a wide orbit with separation $R \gg r$.   We define the relative separation vector of the two-body system and the vector from the center of mass of the two-body system to the third body by
\begin{equation}
{\bm x} \equiv {\bm x}_1 - {\bm x}_2 \,, \quad {\bm X} \equiv {\bm x}_3 - {\bm x}_0  \,,
\label{xdef}
\end{equation}
where
\begin{equation}
{\bm x}_0 \equiv \frac{m_1 {\bm x}_1 + m_2 {\bm x}_2}{m} \,,
\label{x0def}
\end{equation}
where $m \equiv m_1 + m_2$ is the mass of the two-body system.
We work in the center of mass-frame of the entire system, where 
\begin{equation}
m_1 {\bm x}_1 + m_2 {\bm x}_2 + m_3 {\bm x}_3 = m{\bm x}_0 + m_3 {\bm x}_3 = O(c^{-2}) \,,
\label{centerof mass}
\end{equation}
where $O(c^{-2})$ represents a post-Newtonian correction to the center of mass. 
As a result of these definitions, 
\begin{equation}
{\bm x}_1 =  \frac{m_2}{m} {\bm x} - \frac{m_3}{M} {\bm X} ,\,
 {\bm x}_2 = - \frac{m_1}{m} {\bm x} - \frac{m_3}{M} {\bm X} ,\,
 {\bm x}_3 = \frac{m}{M} {\bm X} \,,
 \label{xtransform}
\end{equation}
where $M = m_1 + m_2 + m_3$ is the total mass.  The  $O(c^{-2})$ correction in Eq.\ (\ref{centerof mass})  will not be relevant because only differences between position vectors appear in the equations of motion, and because velocities that are derived from these relations appear in terms that are already of PN order.  We define the velocities ${\bm v} \equiv d{\bm x}/dt$, ${\bm V} \equiv d{\bm X}/dt$, accelerations ${\bm a} \equiv d{\bm v}/dt$, ${\bm A} \equiv d{\bm V}/dt$, distances $r \equiv |{\bm x}|$, $R \equiv
|{\bm X}|$, and unit vectors ${\bm n} \equiv {\bm x}/r$ and ${\bm N} \equiv {\bm X}/R$. For future use we define the symmetric reduced mass $\eta \equiv m_1 m_2/m^2$ and the dimensionless mass difference $\Delta \equiv (m_1 - m_2)/m$.  

The interaction of the two bodies with the third body depends on ${\bm x}_{13}$ and ${\bm x}_{23}$, which we will express as
\begin{eqnarray}
{\bm x}_{13} &=&  -{\bm X} + \alpha_2 {\bm x} = -R \left [{\bm N} - \alpha_2 (r/R) {\bm n}\right ]\,,
\nonumber \\
{\bm x}_{23} &=&  -{\bm X} - \alpha_1 {\bm x} = -R \left [{\bm N} + \alpha_1 (r/R) {\bm n}\right ]\,, 
\label{x13x23}
\end{eqnarray}
where $\alpha_i \equiv m_i/m$; we will use this to expand quantities such as $1/r_{13}$ and $1/r_{23}$ as power series in $r/R$.  The resulting equations of motion for the binary system have the form, 
\begin{eqnarray}
{\bm a} &=& - \frac{Gm{\bm n}}{r^2} - \frac{Gm_3 \,r}{R^3} \left [ {\bm n} -3({\bm n} \cdot {\bm N}) {\bm N} \right ]
+ \frac{1}{c^2}[{\bm a}]_{\rm Binary} 
\nonumber \\
&& \quad 
+ \frac{1}{c^2}[{\bm a}]_{\rm Cross}
+ O\left (\frac{G^2 m m_3 r^{1/2}}{c^2 R^{7/2}}\right ) \,,
\label{acc13}
\end{eqnarray}
where we have expanded the Newtonian term from the third body to quadrupole order, and where the Binary and Cross terms are given by
\begin{eqnarray}
[{\bm a}]_{\rm Binary} &=& \frac{Gm{\bm n}}{r^2} \left [(4+2\eta) \frac{Gm}{r} - (1+3\eta)v^2 + \frac{3}{2} \eta \dot{r}^2 \right ] 
\nonumber \\
&& \qquad
+ (4-2\eta) \frac{Gm\dot{r} {\bm v}}{r^2} \,,
 \label{eqn:abinary}
 \\
\left [{\bm a}\right]_{\rm Cross} &=& \frac{G\mu_3 \Delta}{r^2} \left [ 2{\bm n} ({\bm v} \cdot {\bm V}) 
+ {\bm v}  ({\bm n} \cdot {\bm V}) \right ]
 \nonumber \\
&&
 + \frac{G \mu_3 {\bm n}}{r^2} \left [ \frac{5GM}{R} + \alpha_3 \left ( V^2 + \frac{3}{2} \dot{R}^2 \right ) \right ]
 \nonumber \\
&&
+ \frac{Gm_3 \Delta}{R^2} \left [ \frac{Gm}{2r} \left ( {\bm N} - 9w {\bm n} \right ) + 4 {\bm v} ({\bm v} \cdot {\bm N}) - v^2 {\bm N} \right ]
 \nonumber \\
&&
-\frac{Gm_3}{R^2} \left [ (4-2\alpha_3) {\bm N} ({\bm v} \cdot {\bm V}) -4{\bm V}  ({\bm N} \cdot {\bm v}) -(3+\alpha_3) \dot{R} {\bm v}   \right ]
 \nonumber \\
&&
+\frac{G^2 mm_3}{R^3}\biggl [ (4-\eta)\left ( {\bm n} -3w {\bm N} \right )
-\frac{1}{2} (4 -13\eta) {\bm n} \left ( 1- 3w^2 \right ) \biggr ]
 \nonumber \\
&&
+ \frac{Gm_3 r}{R^3} (1-3\eta) \bigl [ 4{\bf v} \left \{ \dot{r} -3w({\bm v} \cdot {\bm N}\right \} 
- v^2 \left \{ {\bm n} -3w {\bm N} \right \} \bigr]
  \,,
  \label{eqn:across}
\end{eqnarray}
where $\mu_3 \equiv m_3 m/M$, $\alpha_3 \equiv m_3/M$, $\dot{r} \equiv {\bm n} \cdot {\bm v}$, $\dot{R} \equiv {\bm N} \cdot {\bm V}$ and $w \equiv {\bm n} \cdot {\bm N}$.  Although we are nominally considering effects at linear order in $m_3$, the additional factors of $m_3$ that appear via $\alpha_3$ and $\eta_3$ are kinematical in nature, arising from the transformation of velocities from ($v_1$, $v_2$, $v_3$) to ($v$, $V$), and allow us to consider cases where $m_3 \gg m$, as long as $(m_3/m)(a/R)^3 \ll 1$.
Recalling that $v^2 \sim Gm/r$, and $V^2 \sim GM/R$ we see that the six cross terms scale roughly as $(Gmm_3/R^3c^2) \times (R/r)^{n/2}$, where $n = 5,\, 4,\, 2,\, 1,\, 0,\, 0$, respectively.

Treating the third body in an analogous way and defining ${\bm A} \equiv d^2 {\bm X}/dt^2$, we obtain
\begin{equation}
{\bm A} = - \frac{GM{\bm N}}{R^2} + \frac{3}{2}\frac{GM \eta r^2}{R^4} \left [ {\bm N} \left (1- 5w^2 \right) + 2w{\bm n} \right ]
+ O \left (\frac{1}{c^2} \right ) \,.
\label{acc3}
\end{equation}
Explicit expressions for the PN terms will not be needed for this discussion.

\section{Conservation of energy and angular momentum}
\label{conservation}

It is straightforward, to show, either by truncating the full PN expressions for energy and angular momentum of an $N$-body system (see, eg. Paper I, Eq. (3.2a) for the energy), or by constructing conserved quantities directly from the equations of motion (\ref{acc13}) and (\ref{acc3}),  that the conserved total energy and angular momentum are given by
\begin{eqnarray}
E &=& \frac{1}{2} \mu v^2 - \frac{G\mu m}{r} + \frac{1}{2} \mu_3 V^2 - \frac{G\mu_3 M}{R}
+ \frac{1}{2} \frac{G\mu m_3 r^2}{R^3} (1-3w^2)
\nonumber \\
&& \quad   + \frac{1}{c^2} [E]_{\rm Binary}
+ O \left (\frac{m_3}{c^2} \right )\,,
\label{eqn:Econs}
 \\
{\bm L} &=& \mu {\bm x} \times {\bm v} + \mu_3 {\bm X} \times {\bm V} 
+ \frac{1}{c^2} [{\bm L}]_{\rm Binary}
+ O \left (\frac{m_3}{c^2} \right ) \,,
\label{eqn:Lcons}
\end{eqnarray}
where $\mu_3 = m_3 m/M$, and
where the PN contributions from the binary system are given by
\begin{eqnarray}
[E]_{\rm Binary} &=&  \frac{3}{8} \mu (1-3\eta) v^4 + \frac{1}{2} \frac{Gm\mu}{r} \left [ (3+\eta)v^2 + \frac{Gm}{r} + \eta \dot{r}^2 \right ] \,,
\label{eqn:Ebinary}
 \\
\ [{\bm L}]_{\rm Binary}
&=& \mu {\bm x} \times {\bm v} \left [ \frac{1}{2} (1-3\eta) v^2 + (3+\eta) \frac{Gm}{r} \right ] \,,
\label{eqn:Lbinary}
\end{eqnarray}
Explicit forms for the cross-term contributions to $E$ and $\bm L$, of order $m_3/c^2$, will not be needed.

We now consider the simplified problem in which the outer star is on a circular orbit on the $X-Y$ plane.  The inner binary is described by an osculating Keplerian orbit, defined by the equations
\begin{eqnarray}
r &\equiv& a(1-e^2)/(1+e \cos f) \,,
\nonumber \\
{\bm x} &\equiv& r {\bm n} \,,
\nonumber \\
{\bm n} &\equiv& \left [ \cos \Omega \cos(\omega + f) - \cos \iota \sin \Omega \sin (\omega + f) \right ] {\bm e}_X 
\nonumber \\
&&
 + \left [ \sin \Omega \cos (\omega + f) + \cos \iota \cos \Omega \sin(\omega + f) \right ]{\bm e}_Y
\nonumber \\
&&
+ \sin \iota \sin(\omega + f) {\bm e}_Z \,,
\nonumber \\
{\bm \lambda} &\equiv& d{\bm n}/df \,, \quad \hat{\bm h} \equiv {\bm n} \times {\bm \lambda} \,,
\nonumber \\
{\bm h} &\equiv& {\bm x} \times {\bm v} \equiv \sqrt{Gma(1-e^2)} \, \bm{\hat{h}} \,,
\label{keplerorbit}
\end{eqnarray}
where $f$ is the orbital phase or true anomaly and $\Omega$ is the angle of the ascending node.   From the given definitions, it is evident that ${\bm v} = \dot{r} {\bm n} + (h/r) {\bm \lambda}$ and $\dot{r} = (he/p) \sin f$.  The orbit elements $a$, $e$, $\omega$, $\iota$ and $\Omega$ are functions of $f$ when the orbit is not purely Keplerian. 

The outer binary is described by the equations $\bm{X} = R \bm{N}$ and $\bm{V} = \Omega_3 R \bm{\Lambda}$, where $\Omega_3 = (GM/R^3)^{1/2}$, and where
\begin{eqnarray}
\bm{N} &=& \bm{e}_X \cos \Omega_3 t + \bm{e}_Y \sin \Omega_3 t \,,
\nonumber \\
\bm{\Lambda} &=& -\bm{e}_X \sin \Omega_3 t + \bm{e}_Y \cos \Omega_3 t \,,
\nonumber \\
\bm{H} &=& \bm{N} \times \bm{\Lambda} = \bm{e}_Z \,.
\label{eqn:outer}
\end{eqnarray}
We shall ignore perturbations of the third body's orbit due to the binary; these will not be germane to the present discussion.

Retaining only the Newtonian and PN binary terms in the conserved energy and in the $Z$ component of the angular momentum, averaging over an orbit of the third body, and expressing the result in terms of the osculating orbit elements of the inner binary, we obtain
\begin{eqnarray}
E &=& -\frac{G\mu m}{2a}  - \frac{1}{4} \frac{G\mu m_3 a^2}{R^3} (1-e)^2 \left ( 1-3 \sin^2 \iota \, \sin^2 \omega \right )
\nonumber \\
&& 
+ \frac{1}{8} \frac{\mu}{c^2} \left (\frac{Gm}{a} \right )^2 \left [ \frac{4+4(3+\eta)(1+e) + 3(1-3\eta)(1+e)^2 }{(1-e)^2} \right ]
\nonumber \\
&& 
+ O \left ( \frac{m_3}{c^2}  \right ) \,,
\label{eqn:Eaverage}
\\
L_Z &=& \mu [Gma(1-e^2)]^{1/2} \cos \iota \left [ 1 + \frac{1}{2}\frac{Gm}{c^2 a} \frac{(1-3\eta)(1+e)+ 2(3+\eta)}{1-e} \right ]
\nonumber \\
&&  + O \left ( \frac{m_3}{c^2}  \right ) \,.
\label{eqn:Laverage}
\end{eqnarray}
We have ignored the constant contributions to the energy and $L_Z$ from the circular orbit of the third-body alone.  Since $E$ and $L_Z$ are known to be constants, independent of true anomaly $f$,  we have displayed them with all orbit elements evaluated at pericenter, $f=0$. 

The expression for $E$ presents us with a conundrum.  In the standard Newtonian Kozai-Lidov problem, the semi-major axis $a$ suffers {\bf no} secular variations, and the Newtonian secular variations in $e$ and $\iota$ are all of order $(m_3/m)(a/R)^3$, and thus of higher order.   When the PN binary effects are included, they do not contribute additional secular variations in $a$, $e$ and $\iota$.  However, the pericenter $\omega$ is {\em not} constant, but increases via the standard PN secular effect, which we assume dominates other sources of pericenter precession.  But Eq.~(\ref{eqn:Eaverage}) shows that the interaction energy between the binary system and the third body varies with pericenter angle $\omega$.  This makes sense physically: when $\omega =0$, the eccentric binary orbit lies more or less close to the $X-Y$ plane, with the pericenter and apocenter lying in the plane, while when $\omega = \pi/2$, the eccentric orbit extends well above and below the plane of the third body.  It makes sense that the interaction between the binary and the third body should be quite different in the two cases.   But to the order of approximation shown in Eq.\ (\ref{eqn:Eaverage}), and over a pericenter precession timescale, the total energy must be constant, while the orbit elements $a$ and $e$ are also constant.    What has gone wrong?

In the next section we will demonstrate that nothing has gone wrong.  We will show explicitly that $E$ and $L_Z$ are in fact conserved over a pericenter advance timescale, {\em if and only if} one takes into account the contributions of the PN cross terms in the equations of motion to the secular variation of the orbit elements $a$, $e$ and $\iota$.    

\section{Conserved quantities on relativistic precession timescales}
\label{resolution}

In Paper I~\cite{2014PhRvD..89d4043W}, we solved the Lagrange planetary equations for the orbit elements of the inner binary, including the PN cross terms in the equation of motion.  As we emphasized in that paper, it is essential to find the periodic perturbations in the orbit elements induced by the Newtonian third-body perturbation and by the post-Newtonian binary perturbations, and to substitute those periodic effects {\em back} into the planetary equations, because they will induce cross-term effects of the same order as those in the equations of motion.  In addition, in converting the planetary equations from time derivatives to derivatives with respect to true anomaly $f$, it is essential to use the proper conversion
${df}/{dt} = {h}/{r^2}  - \dot{\omega} - \dot{\Omega} \cos \iota$, which can also introduce cross-terms.
We integrated the equations over an orbit of the inner binary and averaged over an orbit of the third body to determine the secular variations in the orbit elements.  The results can be divided into post-Newtonian binary terms, Newtonian terms from the third body, labelled ``K'' for Kozai, and cross terms
(see Paper I, Eqs. (4.13) and (4.14)):
\begin{eqnarray}
\langle \Delta \omega \rangle &=& \langle \Delta \omega \rangle_{\rm Binary} + \langle \Delta \omega \rangle_{\rm K} + \langle \Delta \omega \rangle_{\rm Cross}
\,,
\nonumber \\
\langle \Delta e \rangle &=& \langle \Delta e \rangle_{\rm K} + \langle \Delta e \rangle_{\rm Cross}
\,,
\nonumber \\
\langle \Delta \iota \rangle &=& \langle \Delta \iota \rangle_{\rm K} + \langle \Delta \iota \rangle_{\rm Cross}
\,,
\nonumber \\
\langle \Delta a \rangle &=& \langle \Delta a \rangle_{\rm Cross} \,,
\label{eqn:Deltaelements}
\end{eqnarray}
where $\langle \Delta \omega \rangle_{\rm Binary}$ is the usual post-Newtonian binary pericenter advance, given by 
\begin{equation}
\langle \Delta \omega \rangle_{\rm Binary} = \frac{6\pi Gm}{c^2 a(1-e^2)} \,;
\label{eqn:Deltaomega}
\end{equation}
since we are assuming that this is the dominant contribution to pericenter precession, we will not display the smaller Kozai and cross-term contributions.  Notice that the semi-major axis suffers no secular changes induced by either the PN binary or the Newtonian third-body perturbations, while the eccentricity and inclination suffer no secular changes from the PN binary terms.  The Newtonian Kozai contributions to $\langle \Delta e \rangle$ and $\langle \Delta \iota \rangle$ are given by
\begin{equation}
\langle \Delta e \rangle_{\rm K} = \frac{15\pi}{2} \frac{m_3}{m} \left ( \frac{a}{R} \right )^3 e(1-e^2)^{1/2} \sin^2 \iota \sin \omega \, \cos \omega \,,
\label{delekozai} 
\end{equation}
with $\langle \Delta \iota \rangle_{\rm K} = -e \langle \Delta e \rangle_{\rm K}  \cot \iota /(1-e^2)$.  It is useful to recall that these last relations imply that $\langle \Delta [(1-e^2)^{1/2} \cos \iota ] \rangle_K =0$, expressing the conservation of $L_Z$ at Newtonian order.  The PN cross-term contributions to $\langle \Delta a \rangle$, $\langle \Delta e \rangle$ and $\langle \Delta \iota \rangle$ are given by Paper I, Eqs.\ (4.14): 
\begin{eqnarray}
\langle \Delta a \rangle_{\rm Cross} &=& -\frac{15\pi}{2}\frac{Gm_3}{c^2} \left ( \frac{a}{R} \right )^3   F(e,\eta) \sin^2 \iota \sin 2\omega \,,
\label{delacross}
\\
\langle \Delta e \rangle_{\rm Cross} &=& - \frac{15\pi}{8}\frac{Gm_3}{a c^2} \left ( \frac{a}{R} \right )^3 \biggl \{ G(e,\eta) \sin 2\omega  
\nonumber \\
&& \qquad
- 12\pi \frac{e}{(1-e^2)^{1/2}}\cos 2 \omega  \biggr \} \sin^2 \iota  \,,
\label{delecross}
\\
\langle \Delta \iota \rangle_{\rm Cross} &&=- \frac{15\pi}{8}\frac{Gm_3}{a c^2} \left ( \frac{a}{R} \right )^3 \biggl \{ H(e,\eta) \sin 2\omega 
\nonumber \\
&& \qquad
+ 12\pi \frac{e^2}{(1-e^2)^{3/2}}\cos 2 \omega  \biggr \} \sin \iota \cos \iota \,,
\label{deliotacross}
\end{eqnarray}
where
\begin{eqnarray}
F(e,\eta) &\equiv&  \frac{e(1+e)^2\left [ 7+3e-\eta(3+4e) \right ]}{(1-e)(1-e^2)^{3/2}} 
   +\frac{6}{5}\frac{1-e}{1+e} \,,
 \nonumber \\
 G(e,\eta) &\equiv&  \frac{(1+e)^2 \left [ (3+7e)-(1+6e)\eta -f(e,\eta) \right ]}{(1-e)(1-e^2)^{1/2}} 
\nonumber \\
&& \qquad 
+\frac{4}{5} \frac{(1-e)^2(2+4e-3e^2)}{e^3} \,,
\nonumber \\
H(e,\eta) &\equiv&  \frac{e(1+e)^2 \left [ (3+7e)-(1+6e)\eta + f(e,\eta) \right ]}{(1-e)(1-e^2)^{3/2}} 
\nonumber \\
&& \qquad 
-\frac{8}{5} \frac{(1-e)^3(1+3e)}{e^2(1-e^2)}  \,,
\label{eqn:FGH}
\end{eqnarray}
and
\begin{eqnarray}
f(e,\eta) &\equiv& \frac{1}{5e^3(1+e)} \biggl [ 8-16e-24e^2+109e^3+114e^4+43e^5+16e^6
\nonumber \\
&& \qquad -\eta e^3(15+47e+76e^2+37e^3) \biggr ] \,.
\label{eqn:functionf}
\end{eqnarray} 

These are the secular changes in $a$, $e$ and $\iota$ over one orbit.  Nominally they would grow linearly in time, except for the fact that the angle of pericenter $\omega$ is changing with time at the rate per orbit dominated by Eq.~(\ref{eqn:Deltaomega}).  Thus we can combine this with Eqs.\ (\ref{eqn:Deltaelements}),  to obtain the equation, for each element $Z$, 
\begin{equation}
Z = Z_0 + \int (\Delta Z/\Delta t) dt = Z_0 + \int (\Delta Z/\Delta \omega) d\omega \,. \label{integrate_element}
\end{equation}
Given that $a$, $e$ and $\iota$ are constant to lowest order at this level of approximation, we can carry out the integrations over $\omega$ to obtain
\begin{eqnarray}
a&=& a_0 - \frac{5}{4} a_0 \frac{m_3}{m} \left ( \frac{a_0}{R} \right )^3 F(e_0,\eta)(1-e_0^2)  \sin^2 \iota_0 \, \left (\sin^2 \omega - \sin^2 \omega_0 \right ) \,,
\label{asol}
\\
e &=&e_0 + \frac{5}{8}  \left (\frac{c^2 a_0}{Gm} \right ) \frac{m_3}{m} \left ( \frac{a_0}{R} \right )^3 e_0 (1-e_0^2)^{3/2} \sin^2 \iota_0 \left (\sin^2 \omega - \sin^2 \omega_0 \right )
\nonumber \\
&& - \frac{5}{16}  \frac{m_3}{m} \left ( \frac{a_0}{R} \right )^3 (1-e_0^2)
\biggl \{ G(e_0,\eta) \left (\sin^2 \omega -   \sin^2 \omega_0 \right )
\nonumber \\
&& \qquad
- 6\pi \frac{e_0}{(1-e_0^2)^{1/2}} \left (\sin 2 \omega -   \sin 2 \omega_0 \right )  \biggr \} \sin^2 \iota_0  \,,
\label{esol}
\\
\iota &=& \iota_0 - \frac{5}{8}  \left (\frac{c^2 a_0}{Gm} \right ) \frac{m_3}{m} \left ( \frac{a_0}{R} \right )^3 e_0^2 (1-e_0^2)^{1/2} \sin \iota_0 \cos \iota_0 \left (\sin^2 \omega - \sin^2 \omega_0 \right )
\nonumber \\
&& - \frac{5}{16}  \frac{m_3}{m} \left ( \frac{a_0}{R} \right )^3 (1-e_0^2)
\biggl \{ H(e_0,\eta) \left (\sin^2 \omega -   \sin^2 \omega_0 \right )
\nonumber \\
&& \qquad
+ 6\pi \frac{e_0^2}{(1-e_0^2)^{3/2}} \left (\sin 2 \omega -   \sin 2 \omega_0 \right )  \biggr \} \sin \iota_0 \cos \iota_0 \,,
\label{isol}
\end{eqnarray}
where the subscript $0$ denotes the value of the orbit element at a chosen initial pericenter time.
Notice that, over a pericenter precession timescale, the cross terms induce  {\em Newtonian} level variations in $a$, $e$, and $\iota$, while the Newtonian third-body Kozai perturbations induce a large ``-1PN'' variation in $e$ and $\iota$; these are the terms scaled by the large factor $c^2 a_0/Gm$.   When these results are substituted into Eqs.~(\ref{eqn:Eaverage}) and (\ref{eqn:Laverage}) and expanded to the appropriate order, there is a miraculous cancellation of terms, with all dependence on $\omega$ cancelling, leaving expressions for $E$ and $L_Z$ given by
\begin{eqnarray}
E &=& -\frac{G\mu m}{2a_0}  - \frac{1}{4} \frac{G\mu m_3 a_0^2}{R^3} (1-e_0)^2 \left ( 1-3 \sin^2 \iota_0 \, \sin^2 \omega_0 \right )
\nonumber \\
&& 
+ \frac{1}{8} \frac{\mu}{c^2} \left (\frac{Gm}{a_0} \right )^2 \left [ \frac{4+4(3+\eta)(1+e_0) + 3(1-3\eta)(1+e_0)^2 }{(1-e_0)^2} \right ]
\nonumber \\
&& 
+ O \left ( \frac{m_3}{c^2}  \right ) \,,
\label{Econsfinal}
 \\
L_Z &=& \mu [Gma_0(1-e_0^2)]^{1/2} \cos \iota_0 \left [ 1 + \frac{1}{2}\frac{Gm}{c^2 a_0} \frac{(1-3\eta)(1+e_0)+ 2(3+\eta)}{1-e_0} \right ]
\nonumber \\
&&  + O \left ( \frac{m_3}{c^2}  \right ) \,.
\label{Lconsfinal}
\end{eqnarray}
Since these involve only orbit elements evaluated at the initial pericenter time, they are manifestly constant in time.  The $\sin^2 \omega$ and $\sin 2 \omega$ dependences have disappeared.
For example, in the variation of $a$ in Eq.~(\ref{asol}), the term in $F(e_0,\eta)$ given by $6(1-e_0)/5(1+e_0)$ is exactly what is needed to cancel the $\sin^2 \omega$ term in the energy, leaving $\sin^2 \omega_0$, while the remaining part of $F(e_0,\eta)$ is cancelled by the ``-1PN'' variations in $e$ in Eq.~(\ref{esol}) acting on the PN contribution to the energy.    In $L_Z$, the ``-1PN'' variations in $e$ and $\iota$ in Eqs.~(\ref{esol}) and (\ref{isol}) exactly cancel each other (this is the standard  result for the conservation of $L_Z$ in the Kozai problem), but the Newtonian variations do not; however these in turn are exactly cancelled by the ``-1PN'' variations in $e$ in the post-Newtonian contribution to $L_Z$.

\section{Concluding remarks}
\label{conclusion}

We have shown that the total energy and $Z$ component of total angular momentum of a hierarchical triple system are manifestly conserved to Newtonian order over a relativistic pericenter precession timescale if and only if post-Newtonian cross-term effects in the equations of motion are taken carefully into account.   Future work will explore the implications of PN cross terms in hierarchical triple systems, along two directions.  One is the numerical integration of the orbit evolution equations (\ref{eqn:Deltaelements}) - (\ref{eqn:functionf}) to explore the possible long-term effects of PN cross terms.  Another is to translate the dynamics of hierarchical triples including cross terms into the language of Hamiltonian dynamics and Delaunay variables, which has dominated the literature of the Kozai-Lidov problem, in order to compare and contrast our approach with other work.

\ack
This work was supported in part by the National Science Foundation,
Grant Nos.\ PHY 12--60995 \& 13-06069.   We thank the organizers of 
the ``Al\'ajar Meeting 2013: Stellar dynamics and growth of massive black holes''  for a stimulating venue for discussions of issues related to this work.

\section*{References}

\bibliographystyle{iopart-num}
\bibliography{refs}

\end{document}